\documentclass[a4paper,12pt]{article}

\usepackage{amsmath}

\usepackage{amssymb,epsfig}

\newtheorem{theorem}{Theorem}[section]
\newtheorem{corollary}{Corollary}[section]

\numberwithin{equation}{section}

\begin{document}


\title{ Exact relativistic treatment of stationary
counter-rotating dust disks II\\ Axis, Disk and Limiting Cases
}

\author{
C. Klein, \\ Laboratoire de Gravitation et Cosmologie Relativistes,\\
Universit\'e P.~et M.~Curie, \\ 4, place Jussieu, 75005 Paris,
France\\
}


\date{} 

\maketitle

\begin{abstract}
This is the second in a series of papers on the construction of
explicit solutions to the stationary axisymmetric Einstein equations
which can be interpreted as counter-rotating disks of dust.  We 
discuss the  class of solutions to the
Einstein equations for disks with constant angular velocity and
constant relative density which was constructed in the first part. 
The metric for these spacetimes is given in terms of theta functions
on a Riemann surface of genus 2. We discuss the metric functions
at the axis of symmetry and the disk.   Interesting limiting
cases are  the Newtonian limit, the static limit, 
and the ultra-relativistic limit of the
solution in which the central redshift diverges. 
\end{abstract}


\section{Introduction}\label{sec.1}
The stationary axisymmetric Einstein equations are of great physical 
importance since their solutions can describe the gravitational field 
of stars and galaxies in thermodynamical equilibrium. In this case the 
matter can 
be approximated as an ideal fluid, but the field 
equations in the matter region do not seem to be integrable. The 
vacuum equations are however equivalent to the Ernst equation 
\cite{ernst} which is completely integrable 
\cite{maison,belzak,neuglinear}. If one considers two-dimensional 
matter distributions as disks which are discussed as models for the 
matter in galaxies or in accretion disks around black-holes, the matter 
leads to a boundary value problem for the vacuum equations. A 
solution to this boundary value problem leads to global spacetimes 
with matter distributions which can be physically interpreted.

The first analytic solution for a stationary disk was identified by 
Neugebauer and Meinel \cite{neugebauermeinel1} as belonging to 
Korotkin's \cite{korot1} algebro-geometric solutions to the Ernst 
equation. A systematic study of these solutions in 
\cite{prl,prd2} made it possible in \cite{prl2} to identify a 
class of disk solutions which can be interpreted as being made up of 
counter-rotating dust. In the first paper of this this series 
\cite{prd3} (henceforth referred to as I), the implications of the 
underlying Riemann surface on the 
boundary data taken at a disk were discussed. It was possible to 
construct the solution \cite{prl2} in this way and to identify the 
range of the physical parameters where the solution is globally 
regular except at the disk where the boundary data are prescribed.

\section{Ernst potential and metric}
\label{sec.2.0}
We will briefly summarize results of I where details of the notation
can be found.  We use the Weyl--Lewis--Papapetrou metric (see e.g.\
\cite{exac})
\begin{equation}
\mathrm{ d} s^2 =-e^{2U}(\mathrm{ d} t+a\mathrm{ d} \phi)^2+e^{-2U}
\left(e^{2k}(\mathrm{ d} \rho^2+\mathrm{ d} \zeta^2)+ \rho^2\mathrm{
d} \phi^2\right), \label{vac1}
\end{equation}
where $\rho$ and $\zeta$ are Weyl's canonical coordinates and
$\partial_{t}$ and $\partial_{\phi}$ are the two commuting
asymptotically timelike respectively spacelike Killing vectors.  With
$z=\rho+\mathrm{ i}\zeta$ and the potential $b$ defined by
\begin{equation}
b_{z}=-\frac{\mathrm{ i}}{\rho}e^{4U}a_{z}
\label{vac9},
\end{equation} 
and $b\to 0$ for $z\to\infty$, we define the complex Ernst potential
$f=e^{2U}+\mathrm{ i}b$ which is subject to the Ernst equation
\cite{ernst}
\begin{equation}
f_{z\bar{z}}+\frac{1}{2(z+\bar{z})}(f_{\bar{z}}+f_z)=\frac{2
}{f+\bar{f}} f_z f_{\bar{z}} \label{vac10}\enspace,
\end{equation}
where a bar denotes complex conjugation in ${\mathbb C}$.  The metric
function $k$ follows from
\begin{equation}
k_{z}=2\rho \frac{f_{z}\bar{f}_{z}}{(f+\bar{f})^{2}} \label{2.10a10}.
\end{equation}

In I we have considered disks which can be interpreted as two
counter-rotating components of pressureless matter, so-called dust. 
The surface  energy-momentum tensor of these models is defined on the
hypersurface $\zeta=0$.  It could be written in the form
\begin{equation} 
    S^{\mu\nu}=\sigma_+ u^{\mu}_+ u^{\nu}_+ +\sigma_{-}
u^{\mu}_- u^{\nu}_- \label{2.0},
\end{equation}
where greek indices stand for the $t$, $\rho$ and $\phi$ component and
where $u_{\pm}=(1,0,\pm \Omega)$.   We gave an explicit solution
for disks with constant angular velocity $\Omega$ and constant
relative density
$\gamma=(\sigma_{+}-\sigma_{-})/(\sigma_{+}+\sigma_{-})$ .  This class
of solutions is characterized by two real parameters $\lambda$ and
$\delta$ which are related to $\Omega$ and $\gamma$ and the metric
potential $U_{0}$ at the center of the disk via,
\begin{equation}
\lambda=2\Omega^{2}e^{-2U_{0}}, \quad 
\delta=\frac{1-\gamma^{2}}{\Omega^{2}}.
\label{2.1}
\end{equation}
We put the radius $\rho_{0}$ of the disk equal to 1 unless otherwise
noted.  Since the radius appears only in the combinations
$\rho/\rho_{0}$, $\zeta/\rho_{0}$ and $\Omega \rho_{0}$ in the physical
quantities, it does not have an independent role.  It is always
possible to use it as a natural lengthscale unless
it tends to 0 as in the case of the ultrarelativistic limit of the one
component disk.   

\begin{figure}[htb]
    \centering
    \epsfig{file=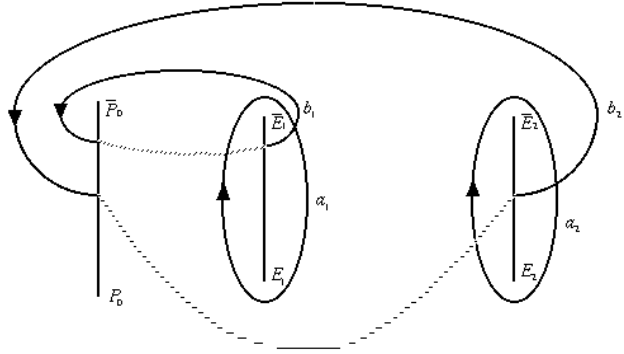,width=10cm}
    \caption{Cut-system.}
    \label{Schnitt1}
\end{figure}

The solution of the Ernst equation we will consider in this paper is
given on a hyperelliptic Riemann surface $\Sigma_{2}$ of genus 2 which
is defined by the algebraic relation
$\mu^{2}(K)=(K+\mathrm{i}z)(K-\mathrm{i}\bar{z})
\prod_{i=1}^{2}(K-E_{i})(K-\bar{E}_{i})$.  We choose
$\mbox{Re}E_{1}<0$, $\mbox{Im}E_{i}<0$ and $E_{1}=-\bar{E}_{2}$ with
$\bar{E}_{2}=\alpha_{1}+\mathrm{i}\beta_{1}$.  We use the  cut-system of 
Fig.~\ref{Schnitt1}. The
base point of the Abel map $\omega$ is $P_{0}=-\mathrm{i}z$. 

The main result of I was the proof of the following
\begin{theorem}
Let $0\leq \delta\leq
\delta_{s}(\lambda):=2\left(1+\sqrt{1+1/\lambda^{2}}\right)$ and $0<\lambda
\leq \lambda_{c}$ where $\lambda_{c}(\gamma)$ is the smallest positive 
value of
$\lambda$ for which $e^{2U_{0}}=0$.  Let
$E_{1}^{2}=\alpha+\mathrm{i}\beta$ with $\alpha$, $\beta$ real and
\begin{equation} \alpha=-1+\frac{\delta}{2},\quad
\beta=\sqrt{\frac{1}{\lambda^{2}}+\delta -\frac{\delta^{2}}{4}}
\label{2.3}.
\end{equation}
Then the solution to the Ernst equation for an energy-momentum tensor 
of the form (\ref{2.0}) with constant
$\Omega$ and constant $\gamma$ can be written in the form
\begin{equation}
f(\rho,\zeta)=\frac{\Theta[m](\omega(\infty^{+})+u)}{
\Theta[m](\omega(\infty^{+})-u)}e^{I}
\label{rel1a},
\end{equation}
where $I= \frac{1}{2\pi\mathrm{ i}}\int\limits_\Gamma \ln
G(\tau)d\omega_{\infty^{+}\infty^-}(\tau)$, where $u_{i}=\frac{1}{2\pi
\mathrm{i}}\int_{\Gamma}^{}\ln G d\omega_{i}$, where $\Gamma$ is the
covering of the imaginary axis in the +-sheet of $\Sigma_{2}$ between
$-\mathrm{i}$ and $\mathrm{i}$, where the characteristic reads
$[m]=\left[
\begin{array}{cc} 
    0 & 0 
    \\ 1 & 1 
\end{array}
\right]$, and where
\begin{equation}
G(\tau)=\frac{\sqrt{(\tau^{2}-\alpha)^2+\beta^2}+\tau^{2}+1}{
\sqrt{(\tau^{2}-\alpha)^2+\beta^2}-(\tau^{2}+1)}.  \label{2.4}
\end{equation}
\end{theorem}

We note that with $\alpha$ and $\beta$ given, the
Riemann surface is completely determined at a given point in the
spacetime, i.e.\ for a given value of $P_{0}$.  
In contrast to algebro-geometric solutions to non-linear evolution 
equations, it depends on the physical coordinates exclusively via the 
branch points   $P_{0}$ and $\bar{P}_{0}$. Since only the modular 
properties of the theta functions are important, the solutions are 
neither periodic or quasiperiodic.

The complete metric (\ref{vac1}) can be expressed via theta functions. 
\begin{theorem}
Let the characteristics $[n_{i}]$ be given by
$        [n_{1}]=\left[ 
	\begin{array}{cc} 
	    0 & 0 \\ 
	    1 & 1 \end{array}
        \right]$ and  $[n_{2}]=\left[ 
	\begin{array}{cc} 
	    0 & 0 \\ 
	    0 & 0
        \end{array} 
	\right]$.
Then the function $e^{2U}$ can
be written in the form
\begin{equation}
e^{2U}=\frac{\Theta[n_{1}](u)\Theta[n_{2}](u)}{
\Theta[n_{1}](0)\Theta[n_{2}](0)}
\frac{\Theta[n_{1}](\omega(\infty^{-}))\Theta[n_{2}](\omega(\infty^{-})
)}{\Theta[n_{1}](\omega(\infty^{-})+u)\Theta[n_{2}](\omega(\infty^{-})+u)}
e^{I} \label{2.5}.
\end{equation}
The metric function $a$ can be expressed via
\begin{equation}
\left(a-a_{0}\right)e^{2U}=-\rho\left( \frac{\Theta[n_{1}](0)
\Theta[n_{2}](0)}{ \Theta[n_{1}](\omega(\infty^{-}))\Theta[n_{2}](
\omega(\infty^{-}))}
\frac{\Theta[n_{1}](u)\Theta[n_{2}](u+2\omega(\infty^{-}))}{
\Theta[n_{1}](u+\omega(\infty^{-}))\Theta[n_{2}](u+\omega(\infty^{-})
)}-1\right) \label{2.6},
\end{equation}
where the constant $a_{0}=-\gamma/\Omega$ is determined by the 
condition that $a$ vanishes on the
regular part of the axis and at infinity.  The metric function
$e^{2k}$ can be put in the form
\begin{equation}
e^{2k}=C\frac{\Theta[n_{1}](u)\Theta[n_{2}](u)}{
\Theta[n_{1}](0)\Theta[n_{2}](0)}\exp\left(
\frac{2}{(4\pi \mathrm{i})^{2}}\int_{\Gamma}^{}\int_{\Gamma}^{} dK_{1}dK_{2}h(K_{1})h(K_{2}) \ln
\frac{\Theta_{o}( \omega(K_{1})-\omega(K_{2}))}{K_{1}-K_{2}}\right)
\label{2.7},
\end{equation}
where $\Theta_{o}$ is a theta function with an odd characteristic,
where $h(\tau)=\partial_{\tau} \ln G(\tau)$, and where $C$ is a
constant which is determined by the condition that $k$ vanishes on the
regular part of the axis and at infinity.
\end{theorem} 

\textbf{Proof:}\\
The metric potential $e^{2U}$ is just the real part of the Ernst
potential.  With the help of Fay's trisecant identity \cite{fay},
$e^{2U}$ was written in \cite{prd2} in the form (\ref{2.5}).  Korotkin
\cite{korot1} gave an expression for the metric function $a$ as a
derivative of theta functions with respect to the argument.  In
\cite{prd2} this formula could be written in the form (\ref{2.6}) free
off derivatives by using the trisecant identity.  The metric function
$e^{2k}$ is related to the so-called $\tau$-function of the linear
system associated to the Ernst equation (see \cite{korotnicolai}). 
This connection made it possible to give the explicit expression for
$k$ in terms of theta functions of (\ref{2.7}) in \cite{korotmat1}.

\section{Axis and branch points}
\label{sec.3}
The axis of symmetry is of physical importance since the multipole
moments such as the Arnowitt-Deser-Misner (ADM) mass and the angular
momentum can be read off from the Ernst potential on the axis.  On the
axis the branch points $P_{0}$ and $\bar{P}_{0}$ coincide which
implies that the Riemann surface becomes singular and 
that some of the periods diverge.  The Ernst
potential is however regular at the axis except at the disk.  The 
proof for this result is based on results by Fay \cite{fay} and Yamada 
\cite{yamada} and was given in
\cite{prd2}. 
We denote here and in
the following the elliptic theta functions with $\vartheta_{i}$ where
$i=1,\ldots,4$ for the characteristics $\left[
\begin{array}{c} 1\\ 1 \end{array} \right]$, $\left[ \begin{array}{c}
1\\ 0 \end{array} \right]$, $\left[ \begin{array}{c} 0\\ 0 \end{array}
\right]$ and $\left[ \begin{array}{c} 0\\ 1 \end{array} \right]$
respectively. If one observes that $a_{0}$ and $C$ are determined by 
the condition that the metric functions  $a$ and $k$ vanish on the 
regular part of the axis, one can summarize the results in 
the following theorem.

\begin{theorem}
We indicate quantities defined on the elliptic Riemann surface
$\Sigma'$ given by $\mu'{}^{2}(\tau)=(\tau^{2}-\alpha)^{2}+\beta^{2}$
with a prime.  Then the Ernst potential on the axis for $\zeta>0$ has
the form
\begin{equation}
f(0,\zeta)=\frac{\vartheta_{4}(\int_{\zeta^{+}}^{\infty^{+}}d\omega'+u_{1}')
-\exp(-\omega_{2}(\infty^{+})-u_{2})
\vartheta_{4}(\int_{\zeta^{-}}^{\infty^{+}}d\omega'+u_{1}')}{
\vartheta_{4}(\int_{\zeta^{+}}^{\infty^{+}}d\omega'-u_{1}')
-\exp(-\omega_{2}(\infty^{+})+u_{2})
\vartheta_{4}(\int_{\zeta^{-}}^{\infty^{+}}d\omega'-u_{1}')}e^{I'+u_{2}}
\label{2.17},
\end{equation} 
where $d\omega_{1}=d\omega'$,
$d\omega_{2}=d\omega'_{\zeta^{-}\zeta^{+}}$, $u_{i}=\frac{1}{2\pi
\mathrm{i}}\int_{\Gamma}^{}\ln G d\omega_{i}$, and where
$I'=\frac{1}{2\pi \mathrm{i}}\int_{\Gamma}^{}\ln G
d\omega'_{\infty^{+}\infty^{-}}$.  The real part of the Ernst
potential can be written in the form
\begin{equation}
e^{2U}=\frac{\vartheta_{4}^{2}(u)}{\vartheta_{4}^{2}(0)}
\frac{\vartheta_{4}^{2}\left(\int_{\zeta^{+}}^{\infty^{-}}d\omega'\right)
-\exp\left(-2\omega_{2}(\infty^{-})\right)
\vartheta_{4}^{2}\left(\int_{\zeta^{-}}^{\infty^{-}}d\omega'\right)}{\vartheta_{4}^{2}\left(u+\int_{\zeta^{+}}^{\infty^{-}}d\omega'\right)
-\exp\left(-2\omega_{2}(\infty^{-})-2u_{2}\right)
\vartheta_{4}^{2}\left(u+\int_{\zeta^{-}}^{\infty^{-}}d\omega'\right)}
\label{2.17a}.
\end{equation}
The constant $a_{0}$ can be expressed via theta functions,
\begin{equation}
a_{0}=\frac{\beta_{1}}{\alpha_{1}}\sqrt{\alpha_{1}^{2}+\beta_{1}^{2}}
\frac{\vartheta_{4}^{4}(0)}{\vartheta_{3}^{2}(\omega(\infty^{-}))
\vartheta_{4}^{2}(\omega(\infty^{-}))}
\frac{\vartheta_{4}(u+2\omega(\infty^{-}))}{\vartheta_{4}(u)}e^{-I'}
\label{2.17b}.
\end{equation}
The constant $C$ is given by
\begin{equation}
1/C=\frac{\vartheta_{4}^{2}(u')}{ \vartheta_{4}^{2}(0)}\exp\left(
\frac{2}{(4\pi \mathrm{i})^{2}}\int_{\Gamma}^{}\int_{\Gamma}^{}dK_{1}dK_{2}h(K_{1})h(K_{2}) \ln
\frac{\vartheta_{1}( \omega'(K_{1})-\omega'(K_{2}))}{K_{1}-K_{2}}
\right) \label{k3}.
\end{equation}

\end{theorem}

The Ernst potential on the axis can be used to determine the 
Ernst potential at the origin, $e^{2U_{0}}$, which is related to
the redshift $z_{R}$ of photons emitted from the center of the disk
and detected at infinity, $z_{R}=e^{-U_{0}}-1$. The
surface $\Sigma'$ admits the involution $\tau\to-\tau$ which implies
$\int_{0^{+}}^{\infty^{+}}d\omega'\equiv\frac{\pi'}{2}, \quad
\int_{0^{-}}^{\infty^{+}}d\omega'\equiv\mathrm{i}\pi $
where $\equiv$ denotes equal up to periods and where $\pi'$ is the
$b$-period on $\Sigma'$.  Similarly one has $u_{1}'=2I'$ and
$u_{2}=\frac{1}{2}\ln G(0) +I'$ where the integral $u_{2}$ is to be
understood as the principal value (this leads to a contribution of
$1/2$ times the residue).  Using $e^{-\omega_{2}(\infty^{+})}=
-\frac{\vartheta_{1}(\int_{0^{+}}^{\infty^{+}}d\omega')}{
\vartheta_{1}(\int_{0^{-}}^{\infty^{+}}d\omega')}$ and
$G(0)=(\sqrt{1+\lambda^{2}}+\lambda)^{2}$ one ends up with
\begin{corollary}
The Ernst potential $f_{0}$ at the center of the disk is given by
\begin{equation}
f_{0}=\frac{(\sqrt{1+\lambda^{2}}-\lambda)X-1}{
\sqrt{1+\lambda^{2}}-\lambda+X} \label{2.18},
\end{equation}
where $X$ is the purely imaginary quantity
\begin{equation}
X=\frac{\vartheta_{3}(u_{1}')\vartheta_{4}(0)}{\vartheta_{1}(u_{1}')
\vartheta_{2}(0)} \label{2.19}.
\end{equation}
\end{corollary}

The ADM mass $M$ and the angular momentum $J$ of the spacetime can be
obtained by expanding the axis potential (\ref{2.17}) in the vicinity
of infinity.  The real part of the Ernst potential for $\epsilon<1$
reads $e^{2U}=1-2M/\zeta +o(1/\zeta)$ and the imaginary part
$b=2J/\zeta^{2}+o(1/\zeta^{2})$ (see e.g.\ \cite{wald}).  We get
\begin{corollary}
The ADM mass is given by the formula
\begin{equation}
M=-D_{\infty^{-}}\ln \vartheta_{4}(u') -\frac{1}{4\pi
\mathrm{i}}\int_{\Gamma}^{} \ln G\, d\omega_{1,\infty^{+}},
\label{disk0c}
\end{equation}
and the angular momentum is given by
\begin{equation} 
    J=
-\frac{\gamma}{\Omega}\left(D_{\infty^{-}}\ln \vartheta_{4}(u')
+D_{\infty^{-}}\ln \vartheta_{2}(u') +\frac{1}{2\pi
\mathrm{i}}\int_{\Gamma}^{}\ln G\, d\omega_{1,\infty^{+}}\right)
\label{disk0d}.  
\end{equation}
Here $D_{P}F(\omega(P))$ denotes the coefficient of the linear term in the
expansion of a function $F$ in the local parameter in the vicinity of 
$P$.
\end{corollary}

Since we concentrate on positive values of $\zeta$, the Riemann
surface can only become singular if $P_{0}$ coincides with
$\bar{P}_{0}$, i.e.\ on the axis, or if it coincides with $E_{2}$. 
Then, as on the axis, several periods of the Riemann surface diverge. 
With the same techniques as on the axis, it was proven in \cite{prd2} 
that the Ernst potential and the metric functions are regular at this 
point.
\begin{theorem}
We denote with a double prime the quantities defined on the Riemann
surface $\Sigma''$ of genus 0 given by
$\mu''{}^{2}(\tau)=(\tau-E_{1})(\tau-\bar{E}_{1})$.  Then the
differentials on $\Sigma_{2}$ reduce for $P_{0}=E_{2}$ to
differentials on $\Sigma''$,
$d\omega_{1}=d\omega''_{E_{2}^{-}E_{2}^{+}}$, $d\omega_{2}=
d\omega''_{\bar{E}_{2}^{-}\bar{E}_{2}^{+}}$ and $I= I''=\frac{1}{2\pi
\mathrm{i}}\int_{\Gamma}^{}\ln G d\omega''_{\infty^{+}\infty^{-}}$. 
The Ernst potential reads
\begin{equation}
f=\frac{\sinh \frac{\omega_{1}(\infty^{+})+u_{1}}{2}}{\sinh
\frac{\omega_{1}(\infty^{+})-u_{1}}{2}} e^{I''} \label{2.17c},
\end{equation}
the function $a$ follows from
\begin{eqnarray} 
    \lefteqn{(a-a_{0}) e^{2U}}&& \nonumber\\
&=&\rho\left(\frac{\sinh \frac{\pi_{12}}{4}}{\sinh
\frac{\omega_{1}(\infty^{+})}{2} \sinh
\frac{\omega_{2}(\infty^{+})}{2}} \times\right.  \\ 
&&\left. 
\frac{\exp\left(\frac{\pi_{12}}{4} \right)
\cosh\frac{u_{1}+u_{2}+2\omega_{1}
(\infty^{+})+2\omega_{2}(\infty^{+})}{2}
-\exp\left(-\frac{\pi_{12}}{4} \right)
\cosh\frac{u_{1}-u_{2}+2\omega_{1}
(\infty^{+})-2\omega_{2}(\infty^{+})}{2}}{ 2\sinh
\frac{u_{1}-\omega_{1}(\infty^{+})}{2} \sinh
\frac{u_{2}-\omega_{2}(\infty^{+})}{2}}-1\right)\nonumber
\label{2.17d}, 
\end{eqnarray}
and the function $e^{2k}$ is given by
\begin{eqnarray}
e^{2k}&=&C \frac{\exp\left(\frac{\pi_{12}}{4} \right)
\cosh\frac{u_{1}+u_{2}}{2}-\exp\left(-\frac{\pi_{12}}{4} \right) \cosh
\frac{u_{1}-u_{2}}{2}}{2\sinh \frac{\pi_{12}}{4}} \nonumber\\
&&\exp\left(\frac{1}{(4\pi \mathrm{i})^{2}}
\int_{\Gamma}^{}\int_{\Gamma}^{}\frac{dK_{1}dK_{2}}{(K_{1}-K_{2})^{2}}
\ln G(K_{1})\ln G(K_{2}) \times \right.\nonumber\\
&&\left.\left(\sqrt{\frac{(K_{1}-E_{1})(K_{2}-\bar{E}_{1})}{
(K_{1}-\bar{E}_{1})(K_{2}-E_{1})}}+\sqrt{\frac{(K_{1}-\bar{E}_{1})(K_{2}-E_{1})}{
(K_{1}-E_{1})(K_{2}-\bar{E}_{1})}}-2\right) \right) \label{2.17e}.
\end{eqnarray}
\end{theorem}

\section{Metric potentials at the disk}\label{sec.4}
In the equatorial plane, the Riemann surface $\Sigma_{2}$ has the 
additional involution $\tau\to-\tau$ which makes it possible to 
express the Ernst potential in terms of elliptic functions (see 
\cite{prd2}).
In I it was shown that there exist algebraic relations between the
real and imaginary parts of the Ernst potential, the function
$Z:=(a-a_{0})e^{2U}$ and their one-sided derivatives at the disk
($\lim_{\zeta \to 0, \zeta>0}$ and $\rho\leq 1$).  The function
$e^{2U}$ itself can be expressed via elliptic theta functions on the 
surface $\Sigma_{w}$ defined by
$\mu_{w}^{2}=(\tau+\rho^{2})((\tau-\alpha)^{2}+\beta^{2})$. We cut the 
surface in a way that the $a$-cut is a closed contour in the upper 
sheet around the cut $[-\rho^{2},\bar{E}]$ and that the $b$-cut starts 
at the cut $[\infty,E]$. The Abel map $w$ is defined for $P\in 
\Sigma_{w}$ as $w(P)=\int_{\infty}^{P}dw$.
\begin{theorem}
Let 
$u_{w}=\frac{1}{\mathrm{i}\pi}\int_{-\rho^{2}}^{-1}\ln G(\sqrt{\tau})dw(\tau)$. 
Then the real part of the Ernst potential at the disk is given by
\begin{eqnarray}
e^{2U}&=&\frac{1}{Y-\delta}\left(-\frac{1}{\lambda}-\frac{Y}{\delta}
\left(\frac{\frac{1}{\lambda^{2}}+\delta}{
\sqrt{\frac{1}{\lambda^{2}}+\delta\rho^{2}}}-\frac{1}{\lambda}\right)
\right.  \nonumber\\ &&\left. 
+\sqrt{\frac{Y^{2}((\rho^{2}+\alpha)^{2}+\beta^{2})}{
\frac{1}{\lambda^{2}}+\delta\rho^{2}}-2Y(\rho^{2}+\alpha) +
\frac{1}{\lambda^{2}}+\delta\rho^{2}}\right) \label{y1},
\end{eqnarray}
where
\begin{equation}
Y=\frac{\frac{1}{\lambda^{2}}+\delta\rho^{2}}{\sqrt{(\rho^{2}+\alpha)^{2}
+\beta^{2}}}\frac{\vartheta_{3}^{2}(u_{w})}{\vartheta_{1}^{2}(u_{w})}
\label{y2}.
\end{equation}
At the disk, the relations
\begin{equation}
\frac{\delta^{2}}{2}(e^{4U}+b^{2})= \left(\frac{1}{\lambda}-\delta
e^{2U}\right)\left(\frac{\frac{1}{\lambda^{2}
}+\delta}{\sqrt{\frac{1}{\lambda^{2}}+\delta \rho^{2}}}-
\frac{1}{\lambda}\right)+\delta\left(\frac{\delta+\rho^{2}}{2}-1\right)
\label{2.9},
\end{equation}
for $e^{2U}$ and $b$,
\begin{equation}
Z^{2}-\rho^{2}+\delta e^{4U}=\frac{2}{\lambda}e^{2U} \label{2.10a}
\end{equation}
for $Z$ and $e^{2U}$, and
\begin{equation}
\left( e^{2U}\right)_{\zeta}=\frac{Z^{2}+\rho^{2}+\delta e^{4U}}{2Z
\rho}b_{\rho} \label{2.10b}
\end{equation}
for the derivatives are valid.
\end{theorem}

\textbf{Proof:}\\
One can define on $\Sigma_{w}$ the divisor $W$ as the solution of the
Jacobi inversion problem
\begin{equation}
\int_{\infty}^{W}dw:=u_{w}.  \label{y3}
\end{equation}
Thus $W$ can be expressed in standard manner via theta functions on
$\Sigma_{w}$, 
\begin{equation}
W+\rho^{2}=\sqrt{(\rho^{2}+\alpha)^{2}+\beta^{2}}
\frac{\vartheta_{3}^{2}(u_{w})}{\vartheta_{1}^{2}(u_{w})} \label{y4}.
\end{equation}
In I it was shown that $W$ is related to $e^{2U}$,
\begin{equation}
W+\rho^{2}=\frac{\delta^{2}((\rho^{2}+\alpha)^{2}+\beta^{2})}{2
\left(\frac{1}{\lambda^{2}}+\delta \rho^{2}\right)}
\frac{\rho^{2}+\frac{2}{\lambda} e^{2U}-\delta e^{4U}}{
\left(\frac{1}{\lambda}-\delta e^{2U}\right)
\left(\frac{\frac{1}{\lambda^{2}}+\delta}{
\sqrt{\frac{1}{\lambda^{2}}+\delta\rho^{2}}}-\frac{1}{\lambda}\right)
+\frac{\delta}{2}(\rho^{2}+2\alpha)-\frac{\delta^{2}}{2}e^{4U}}
\label{y5}.
\end{equation}
Entering with this relation in (\ref{y4}) and solving for $e^{2U}$,
one ends up after some algebraic manipulation with (\ref{y1}).  The
relations (\ref{2.9}) to (\ref{2.10b}) were given in I. This completes
the proof.

To discuss physical quantities in the disk, the behaviour of the
metric in the vicinity of the center and of the rim 
of the disk is important.  We have
\begin{corollary}
At the disk ($\zeta\to 0$, $\zeta>0$), the metric functions and their
derivatives have for $\rho\to 0$ and $\lambda<\lambda_{c}$, 
$\delta\leq \delta_{s}(\lambda)$ an expansion of the
form $F=F_{0} +F_{2}\rho^{2}+F_{4}\rho^{4}+\ldots$ where $F$ stands
for $e^{2U}$, $Z$, $b$, $\left(e^{2U}\right)_{\zeta}$, and $e^{2k}$,
and where the $F_{i}$ are constants with
$b_{0}^{2}=1-4\Omega^{2}-e^{4U_{0}}$ and $Z_{0}=(\gamma/\Omega )
e^{2U_{0}}$.\\
For $\delta=0$ and $\lambda=\lambda_{c}$ the metric functions and 
their derivatives have an expansion of the form
$e^{2U} =  -\frac{\lambda}{2}\rho^{2}+\rho^{4} y_{4}+\ldots$, 
$Z  =  \rho^{2} Z_{2}+\ldots$,
$b  = 
-1+\rho^{4}\lambda y_{4}+\ldots$, 
$\left(e^{2U}\right)_{\zeta}  =  -\sqrt{2\lambda y_{4}}\rho^{2}+
\ldots$,and  $k  = \ln \rho +k_{0}+k_{2}\rho^{2}+\ldots$.\\
For $\delta\neq 0$, the critical value $\lambda_{c}\to \infty$. 
In this case the expansion of the 
metric functions is as for $\delta=0$, $\lambda=\lambda_{c}$ 
or of the form  
$e^{2U}  =  y_{1}\rho +\rho^{3}y_{3}+\ldots$,  
$Z  = 
Z_{1} \rho +\rho^{3}Z_{3}+\ldots$, 
$b  = 
b_{0}+\rho^{2}b_{2}+\ldots$,
$\left(e^{2U}\right)_{\zeta} 
=  s_{1}\rho+\ldots$, 
$k  = \frac{1}{4}\ln \rho
+k_{0}+k_{2}\rho^{2}+\ldots$.\\
At the rim of the disk ($\rho=1$),  the
imaginary part of the Ernst potential vanishes for $\delta>\delta_{s}$ 
and $\rho\leq 1$
as $(1-\rho^{2})^{\frac{3}{2}}$.
\end{corollary}

\textbf{Proof:}\\
Since the Riemann
surface $\Sigma_{w}$ is regular for $0\leq \rho \leq 1$, all
periods are functions of $\rho^{2}$.  The integral $u_{w}$ is also a
smooth function of $\rho^{2}$ which can be seen from
\begin{equation}
\frac{1}{\mathrm{i}\pi}\int_{-\rho^{2}}^{-1}\frac{\ln G(\sqrt{\tau})
d\tau}{\sqrt{(\tau+\rho^{2}) ((\tau-\alpha)^{2}+\beta^{2})}}=
\frac{2}{\pi}\sqrt{1-\rho^{2}}\int_{0}^{1}\tilde{G}((1-\rho^{2})t
+\rho^{2}) dt, \label{y7}
\end{equation}
where
\begin{equation}
\tilde{G}(\tau)=\frac{2\ln
(\sqrt{(\tau+\alpha)^{2}+\beta^{2}}+1-\tau)-\ln
\left(\frac{1}{\lambda^{2}}+\delta \tau\right)}{
\sqrt{(\tau+\alpha)^{2}+\beta^{2}}} \label{2.10a9}.
\end{equation}
Thus the quantity $Y$ in (\ref{y2}) is a smooth function in
$\rho^{2}$.  For $\lambda<\lambda_{c}$ it has the form
$Y(\rho)=Y_{0}+Y_{2}\rho^{2}+\ldots$ This implies via the relations of
Theorem 4.1 that $e^{2U}$, $Z$, $b$ and $\left(e^{2U}\right)_{\zeta}$
are also smooth functions in $\rho^{2}$.  The behaviour of $k$ is
determined by using the relation (\ref{2.10a10}) together with the
condition that $k$ vanishes on the axis.\\
In the case $\gamma=1$ the ultra-relativistic limit $e^{2U_{0}}=1$ is
reached for $\vartheta_{3}(u_{w})=0$ for $\rho=0$ i.e.\ $Y=0$ at the
center of the disk.  Since the theta functions in (\ref{y2}) are
functions in $\rho^{2}$ and since $Y$ is quadratic in the theta
functions, $Y$ is of the form $Y=Y_{4}\rho^{4}+ \ldots$.  Expanding
$e^{2U}$ in (\ref{y1}) for small $Y$ one gets
$e^{2U}=-\frac{\lambda}{2}\rho^{2}+\frac{\lambda}{2}Y $.
For $\lambda\to \infty$ and $0<\gamma<1$, equation
(\ref{y1}) leads to $e^{2U}=\rho y_{1}+\rho^{3}y_{3}+\ldots$ for 
$y_{1}\neq0$ or $e^{2U}=\rho^{2} y_{2}+\rho^{4}y_{4}+\ldots$ for 
$y_{1}=0$. 
For $y_{1}\neq0$ relation (\ref{2.9}) can either be satisfied by a
function $b$ of the form $b=b_{0}+b_{2} \rho^{2}+\ldots$ or
$b=b_{1}\rho+\ldots$.  To decide one has to consider $b_{0}$ which has
the form
$\frac{\delta}{2}b_{0}^{2}=\alpha-\theta+\sqrt{\theta^{2}-2\alpha
\theta +1}$, where
$\theta=\vartheta_{2}^{2}(u_{w})/\vartheta_{1}^{2}(u_{w})$ for
$\rho=0$.  Thus $b_{0}$ can only vanish for $\delta\neq 0$ in this 
case if
$\delta=4$, the static limit, where it vanishes identically.  \\
The behaviour of $b$ at the rim of the disk follows
from (\ref{2.9}) which implies with (\ref{y1})
\begin{eqnarray}
b^{2}&=&\frac{2}{(Y-\delta)^{2}}\left(\delta \alpha-
\frac{1}{\lambda^{2}} -Y\left(\alpha+\frac{2-\rho^{2}
+\delta\lambda^{2}}{1+\lambda^{2}\delta \rho^{2}}\right)+ \right. 
\nonumber \\
&&\left.\frac{\frac{1}{\lambda^{2}}+\delta}{\sqrt{\frac{1}{\lambda^{2}}+\delta
\rho^{2}}}\sqrt{\frac{Y^{2}((\rho^{2}+\alpha)^{2}+\beta^{2})}{
\frac{1}{\lambda^{2}}+\delta\rho^{2}}-2Y(\rho^{2}+\alpha) +
\frac{1}{\lambda^{2}}+\delta\rho^{2}}\right) \label{y11}.
\end{eqnarray}
Since the Riemann surface $\Sigma_{w}$ and therefore its periods are
regular at $\rho=1$, the integral $u_{w}$ is dominated by the integral
in (\ref{y7}) which is proportional to $(1-\rho^{2})^{\frac{3}{2}}$
for $\rho\approx 1$ in the disk.  The theta function
$\vartheta_{1}(x)$ has zeros of first order at $x=0$ which implies
that $Y$ diverges for $\rho\to 1$.  Consequently for $\rho \approx 1$ 
we have
$b^{2}\approx
\frac{2}{Y}\left(\sqrt{\frac{1}{\lambda^{2}}+\delta}-\frac{\delta}{2}\right)
$which implies that $b\propto \vartheta_{1}(u_{w})$ i.e.\
$b\propto(1-\rho^{2})^{\frac{3}{2}}$ in the non-static case.

\section{Limiting cases}\label{subsec.3.3}
The one-component disk which was studied by Bardeen and Wagoner
\cite{bawa} is obtained by simply putting $\delta=0$ in the Ernst
potential (\ref{rel1a}).  This gives the solution of Neugebauer and
Meinel \cite{neugebauermeinel1} in the notation of \cite{prl,prd2}.

\subsection{Newtonian limit}
The Newtonian limit is reached for $\lambda\to0$ for arbitrary $\delta$ which
follows from (\ref{2.1}): If $e^{-U_{0}}-1\ll 1$ and
$\Omega=\Omega \rho_{0}\ll 1$, this means that both the central
redshift and the maximal velocity in the disk (compared to the
velocity of light which is 1 in the used units) are small, which just
defines the Newtonian regime.  We get
\begin{theorem}
In the limit $\lambda\ll 1$, the Ernst potential (\ref{rel1a}) becomes
real with $U$ given by
\begin{equation}
U(\rho,\zeta)=-\frac{1}{4\pi
\mathrm{i}}\int_{-\mathrm{i}}^{\mathrm{i}}
\frac{2\lambda(\tau^{2}+1)}{\sqrt{(\tau-\zeta)^{2}+\rho^{2}}}d\tau
\label{2.11}.
\end{equation}
The metric functions $a$ and $k$ are of order $\Omega^{3}$ and
$\Omega^{4}$ respectively.
\end{theorem}

Equation (\ref{2.11}) just describes the Maclaurin disk in the
notation of I. The metric has the behaviour one expects from a general
post-Newtonian expansion.\\
\textbf{Proof:}\\
In the limit $\lambda\to0$, the branch points $E_{i}$ and
$\bar{E}_{i}$, $i=1,2$, tend to infinity.  To treat this limiting
case, we use the conformal transformation $\tau\to
\tau/\sqrt{\lambda}$.  On the
transformed surface, the cut
$[\sqrt{\lambda}P_{0},\sqrt{\lambda}\bar{P}_{0}]$ collapses as on the 
axis, whereas
the remaining cuts remain finite.  In the limit
$\lambda\to 0$ the Abelian integrals can be expressed in terms of
quantities on the elliptic Riemann surface $\tilde{\Sigma}$ given by
$\tilde{\mu}{}^{2}(\tau)=\tau^{4}+1$. We get
\begin{equation}
f=e^{I-u_{1}-u_{2}}\frac{\vartheta_{3}(u_{1})\vartheta_{4}(0)-
\vartheta_{2}(0)\vartheta_{1}(u_{1})\exp\left(u_{2}+\frac{u_{1}}{2}\right)}{
\vartheta_{3}(u_{1})\vartheta_{4}(0)+
\vartheta_{2}(0)\vartheta_{1}(u_{1})\exp\left(-u_{2}-\frac{u_{1}}{2}
\right)} \label{2.20a}.
\end{equation}

The $u_{i}$ and $I$ are however not Abelian integrals.  We have
$\oint_{a_{1}}\frac{\tau^{n}d\tau}{\mu(\tau)}\propto
\lambda^{1-\frac{n}{2}} $,
whereas the corresponding $a_{2}$-periods are proportional to
$\lambda$.  With $\ln G\approx 2\lambda(\tau^{2}+1)$ we thus find that
$u_{1}$ and $I$ are proportional to $\lambda^{\frac{3}{2}}$ and
$u_{2}=\frac{1}{2\pi \mathrm{i}}\int_{\Gamma}^{}
\frac{2\lambda(\tau^{2}+1)d\tau}{\sqrt{(\tau-\zeta)^{2}+\rho^{2}}}$ in
lowest order.

The leading contributions in (\ref{2.20a}) are consequently given by
$u_{2}$ in
$f=1-u_{2} $.
Since $f=e^{2U}$ in this case, we get (\ref{2.11}).  In a similar way
the lowest order contributions to the imaginary part can be obtained
from (\ref{2.20a}).  They arise from the term $\vartheta_{1}(u_{1})$
which is an odd and purely imaginary function.  It has zeros of first
order which implies that $b\propto u_{1}$ which is of order
$\lambda^{\frac{3}{2}}$ i.e.\ of order $\Omega^{3}$.  Thus $a$ is also
of order $\Omega^{3}$.
For the metric function $k$ we get with (\ref{2.7}) as above that the
leading contributions arise from the integral in the exponent which
are quadratic in $G$ and thus of order $\Omega^{4}$.  This completes
the proof.

\subsection{Static limit}
The static limit of the Ernst potential
(\ref{rel1a}) is obtained for $\beta=0$, i.e.\
$\delta_{s}(\lambda)=2(1+\sqrt{1+1/\lambda^{2}})$.
We get 
\begin{theorem}
The function (\ref{rel1a}) becomes real in the limit
$\delta=\delta_{s}(\lambda)$ with the potential $U$ given by
\begin{equation}
    U(\rho,\zeta)=-\frac{1}{4\pi
\mathrm{i}}\int_{-\mathrm{i}}^{\mathrm{i}}
\ln G
\frac{ d\tau}{\sqrt{(\tau-\zeta)^{2}+\rho^{2}}},
\quad G=\left(1-\frac{4}{\delta}(\tau^{2}+1)\right)
\label{2.15}.
\end{equation}
The metric function $a$ vanishes identically whereas the function $k$
is given by
\begin{eqnarray}
k&=&\frac{1}{2(4\pi \mathrm{i})^{2}}\int_{\Gamma}^{}\int_{\Gamma}^{}
dK_{1}dK_{2}\ln G(K_{1})\ln G(K_{2})\nonumber\\
&&\frac{1}{(K_{1}-K_{2})^{2}}\left(
\sqrt{\frac{(K_{1}-i\bar{z})(K_{2}+iz)}{(K_{2}-i\bar{z})(K_{1}+iz)}}+
\sqrt{\frac{(K_{2}-i\bar{z})(K_{1}+iz)}{(K_{1}-i\bar{z})(K_{2}+iz)}}-2\right)
\label{2.15a}.
\end{eqnarray}
\end{theorem}

This is the static disk of Morgan and Morgan \cite{morgan} with
uniform rotation.

\textbf{Proof:}\\
In the limit $\delta\to \delta_{s}$, the branch points $E_{i}$ and
$\bar{E}_{i}$ coincide for $i=1,2$ on the real axis since
$\delta_{s}\geq 4$ and thus $\alpha(\delta_{s}) =\delta_{s}/2-1>0$. 
Mathematically this limit corresponds to the standard solitonic limit
of algebro-geometric solutions of evolution equations, see e.g.\
\cite{mumford2}.  In the cut-system Fig.~\ref{Schnitt1} the $b$-periods
$\pi_{11}$ and $\pi_{22}$ diverge whereas the remaining periods are
finite.  Thus in this limit the theta functions in (\ref{rel1a})
become identical to 1 whereas the differential of the third kind
$d\omega_{\infty^{+}\infty^{-}}\to
-\frac{d\tau}{\sqrt{(\tau-\zeta)^{2}+\rho^{2}}}$.  The limit of
(\ref{2.4}) for $\delta=\delta_{s}$ is (\ref{2.15}) which is identical
to
$\ln G=4U_0+\ln \left(1+2\lambda e^{-2U_{0}}\tau^2\right)
$,
i.e.\ the form which was given for the Morgan and Morgan disk with
constant $\Omega$ in \cite{cqg}.  Since the theta functions in
(\ref{2.6}) all tend to one, we find that $a$ is identical to zero in
this case with $a_{0}=-\gamma/\Omega$ being equal to zero for
$\gamma=0$.  In the expression for the metric function $k$ (\ref{2.7})
the theta functions again tend to 1.  To evaluate the integral in the
exponent, we integrate by parts with respect to both $K_{1}$ and
$K_{2}$ using the fact that $\ln G$ vanishes at the limits of
integration.  The differential of the second kind then takes the form
given in (\ref{2.15a}).  Since $k$ obviously vanishes on the axis, the
constant $C$ is identical to 1 in this limit.  This completes the
proof.

\subsection{Ultra-relativistic limit}\label{sec.5}
The ultra-relativistic limit is reached if the redshift of photons 
emitted at the center of the disk and detected at infinity diverges 
which is equivalent to the fact that photons from the center of the 
disk cannot escape to infinity as from a horizon of a 
black-hole. In the case $\gamma=1$ the limit is reached
for $\vartheta_{4}(u')=0$.  This implies with (\ref{2.17b}) and
(\ref{k3}) that both constants $a_{0}$ and $C$ diverge as
$\epsilon\to1$. 
The axis is in fact singular in the sense that the metric function
$e^{2U}$ vanishes there identically which can be seen from
(\ref{2.17a}).  The Ernst potential is identical to $-\mathrm{i}$ on
the axis for $\zeta>0$.   A
consequence of the diverging constant $a_{0}$ is that the angular
velocity $\Omega=\Omega \rho_{0}$, which is the coordinate angular velocity in the
disk as measured from infinity, vanishes.  
Since this implies that either $\rho_{0}$ or $\Omega$ vanish,
Bardeen and Wagoner \cite{bawa} argued that
the spacetime can be interpreted in the limit $\epsilon\to 1$ and
$\rho_{0}\to 0$ as the extreme Kerr metric in the exterior of the
disk.  In \cite{prl} it was shown that such a limit (diverging
multipoles, singular axis,\ldots) can occur in general hyperelliptic
solutions and can always be interpreted as an extreme Kerr spacetime. 
For an algebraic treatment of the ultra-relativistic limit of the
Bardeen-Wagoner disk see \cite{mexico}.  We have 
\begin{theorem}
Let $f$ be an Ernst potential of the form (\ref{rel1a}), which is
regular except at the disk, with $\vartheta_{4}(u')=0$ and
$\lim_{\rho_{0}\to 0} \rho_{0}a_{0}=:-2m$ where $m$ is a finite
non-zero positive real constant.  Then in the limit $\rho_{0}\to 0$,
the Ernst potential (\ref{rel1a}) for $\rho^{2}+\zeta^{2}\neq0$ takes
the form
\begin{equation}
f=\frac{\rho^{2}+\zeta^{2}+m(\mathrm{i\zeta
-\sqrt{\rho^{2}+\zeta^{2}}})}{ \rho^{2}+\zeta^{2}+m(\mathrm{i\zeta
+\sqrt{\rho^{2}+\zeta^{2}}})} \label{ultra6}.
\end{equation}
\end{theorem}

In the ultra-relativistic limit of the above disks for $\gamma=1$,
this implies that the spacetime becomes an extreme Kerr spacetime with
$m=\frac{1}{2\Omega}$.\\
\textbf{Proof:}\\
The most direct way to prove this statement seems to determine the
potential in the limit on the axis and then to use a theorem of Hauser
and Ernst \cite{hauser} that an Ernst potential is uniquely determined
for given sources if it is given on some finite regular part of the
axis.  The potential on the axis is given by (\ref{2.17}).  The limit
$\rho_{0}\to 0$ is equivalent to the limit $\zeta\to \infty$ as used
in the calculation of the ADM-mass and the angular momentum in section
3 with $\vartheta_{4}(u')=0$.  The constant $a_{0}$
can be written with the help of Fay's trisecant identity \cite{fay} as
\begin{equation}
    a_{0}=-2\mathrm{i}\frac{D_{\infty^{-}}\vartheta_{1}(0)
\vartheta_{4}(u'+2\omega(\infty^{-}))}{\vartheta_{1}(2\omega(\infty^{-}))
\vartheta_{4}(u')}
    \label{a0},
\end{equation}

This implies with identities between elliptic theta functions 
(see e.g.\ \cite{algebro}) and 
the fact that $u'\equiv \pi'/2$, the $b$-period on $\Sigma'$, 
that
$a_{0}=
-2D_{\infty^{-}}\ln \vartheta_{4}(u') $.  
Expanding the axis potential (\ref{2.17}) in first order of
$\rho_{0}$, one thus gets
\begin{equation}
f(0,\zeta)=\frac{\zeta+m(\mathrm{i}-1)}{\zeta+m(\mathrm{i}+1)}
\label{ultra12},
\end{equation}
which is the potential (\ref{ultra6}) on the axis.  This completes the
proof.

For $\gamma<1$ the axis
remains regular except at the origin, 
the constants $a_{0}$ and $C$ in (\ref{2.17b}) and
(\ref{k3}) remain finite here
since they can only diverge if $\vartheta_{4}(u')=0$ which can happen
only for $\gamma=1$.  The integrals in the respective exponents of
(\ref{2.17b}) and (\ref{k3}) are always finite though $\ln G(\tau)$
has a term $\ln \tau$ in the limit $\lambda\to \infty$ as can be
easily seen.  The ultrarelativistic 
limit of the disks with counter-rotation is thus a disk of finite 
radius with diverging central redshift.

\noindent \emph{ Acknowledgment}\\
I thank  R.~Kerner, D.~Korotkin, H.~Pfister, O.~Richter and
U.~Schaudt for helpful remarks and hints.  This work was 
supported by the Marie-Curie program of the European Community.

\end{document}